\documentclass[5p,twocolumn,times,number]{elsarticle}

\usepackage{lineno}

\usepackage{graphicx}
\usepackage{amsmath}   
\usepackage{url}
\usepackage{float}
\def\pmbanner{}

\begin{document}

\begin{frontmatter}

\title{\pmbanner HERMES: An ultra-wide band X and gamma-ray transient monitor on board a nano-satellite constellation}

\author[add1,add2]	{F.~Fuschino\corref{cor}} \ead{fuschino@iasfbo.inaf.it}
\cortext[cor]{Corresponding author}
\author[add1,add2]	{R.~Campana}
\author[add1,add2]	{C.~Labanti}
\author[add3,add4]	{Y.~Evangelista}
\author[add3,add4]	{M.~Feroci}
\author[add7]		{L.~Burderi}
\author[add8]		{F.~Fiore}
\author[add3]		{F.~Ambrosino}
\author[add5,add2]	{G.~Baldazzi}
\author[add16]		{P.~Bellutti}
\author[add17]		{R.~Bertacin}
\author[add12,add13]	{G.~Bertuccio}
\author[add16]		{G.~Borghi}
\author[add9,add10]	{D.~Cirrincione}
\author[add9,add20]	{D.~Cauz}
\author[add21]		{T.~Di~Salvo}
\author[add16]		{F.~Ficorella}
\author[add6]		{M.~Fiorini}
\author[add21]		{A.~Gambino}
\author[add12,add13]	{M.~Gandola}
\author[add14]		{M.~Grassi}
\author[add19]		{A.~Guzman}
\author[add21]		{R.~Iaria}
\author[add18]		{G.~La~Rosa}
\author[add15]		{M.~Lavagna}
\author[add15]		{P.~Lunghi}
\author[add14]		{P.~Malcovati}
\author[add7]		{A.~Maselli}
\author[add20]		{M.~T.~Menna}
\author[add1]		{G.~Morgante}
\author[add17]		{B.~Negri}
\author[add20]		{G.~Pauletta}
\author[add20]		{A.~Papitto}
\author[add3]		{R.~Piazzolla}
\author[add16]		{A.~Picciotto}
\author[add17]		{S.~Pirrotta}
\author[add19]		{S.~Pliego-Caballero}
\author[add17]		{S.~Puccetti}
\author[add10]		{A.~Rachevski}
\author[add11]		{I.~Rashevskaya} 
\author[add7]		{A.~Riggio}
\author[add5,add2]	{L.~Rignanese}
\author[add17]		{M.~Salatti}
\author[add7]		{A.~Sanna}
\author[add19]		{A.~Santangelo}
\author[add15]		{S.~Silvestrini}
\author[add18]		{G.~Sottile}
\author[add19]		{C.~Tenzer}
\author[add9,add10]	{A.~Vacchi}
\author[add10]		{G.~Zampa}
\author[add10]		{N.~Zampa}
\author[add16]		{N.~Zorzi}

\address[add1]{INAF-OAS Bologna, Via Gobetti 101, I-40129 Bologna, Italy}
\address[add2]{INFN sez. Bologna, Viale Berti-Pichat 6/2, I-40127 Bologna, Italy}
\address[add3]{INAF-IAPS,Via del Fosso del Cavaliere 100, I-00133 Rome, Italy}
\address[add4]{INFN sez. Roma 2, Via della Ricerca Scientifica 1, I-00133 Rome, Italy}
\address[add5]{University of Bologna, Dep. of Physics and Astronomy - DIFA, viale Berti Pichat 6/2, I-40127 Bologna, Italy}
\address[add6]{INAF-IASF Milano, Via Bassini 15, I-20100 Milano, Italy}
\address[add7]{University of Cagliari, Dep. of Physics, S.P. Monserrato-Sestu Km 0,700, I-09042 Monserrato (CA), Italy}
\address[add8]{INAF-OATs Via G.B. Tiepolo, 11, I-34143 Trieste}
\address[add9]{University of Udine, Via delle Scienze 206, I-33100 Udine, Italy }
\address[add20]{INFN Udine, Via delle Scienze 206, I-33100 Udine, Italy }
\address[add10]{INFN sez. Trieste, Padriciano 99, I-34127 Trieste, Italy}
\address[add11]{TIFPA-INFN, Via Sommarive 14, I-38123 Trento, Italy}
\address[add12]{Politecnico di Milano, Department of Electronics, Information and Bioengineering, Via Anzani 42, I-22100 Como, Italy}
\address[add13]{INFN sez. Milano, Via Celoria 16, I-20133 Milano, Italy}
\address[add14]{University of Pavia, Department of Electrical, Computer, and Biomedical Engineering, and INFN Sez. Pavia, Via Ferrata 3, I-27100 Pavia, Italy }
\address[add15]{Politecnico of Milano, Bovisa Campus, Via La Masa 34 - I-20156 Milano, Italy}
\address[add16]{Fondazione Bruno Kessler – FBK, Via Sommarive 18, I-38123 Trento, Italy}
\address[add17]{Italian Space Agency - ASI, Via del Politecnico snc, 00133 Roma, Italy}
\address[add18]{INAF/IASF Palermo, Via Ugo La Malfa 153, I-90146 Palermo, Italy}
\address[add19]{University of Tubingen-IAAT, Sand 1, D-72076 Tübingen, Germany}
\address[add20]{INAF-OAR, Via Frascati 33, I-00040, Monte Porzio Catone, Roma, Italy}
\address[add21]{University of Palermo, Dept. of Physics and Chemistry, via Archirafi 36, I-90123, Palermo, Italy}

\begin{abstract}
The High Energy Modular Ensemble of Satellites (HERMES) project is aimed to realize a modular 
X/gamma-ray monitor for transient events, to be placed on-board of a nano-satellite bus (e.g. CubeSat). 
This expandable platform will achieve a significant impact on Gamma Ray Burst (GRB) science and on the 
detection of Gravitational Wave (GW) electromagnetic counterparts: the recent LIGO/VIRGO discoveries 
demonstrated that the high-energy transient sky is still a field of extreme interest.  
The very complex temporal variability of GRBs (experimentally verified up to the millisecond scale) 
combined with the spatial and temporal coincidence between GWs and their electromagnetic counterparts 
suggest that upcoming instruments require sub-microsecond time resolution combined with a transient 
localization accuracy lower than a degree.
The current phase of the ongoing HERMES project is focused on the realization of a technological 
pathfinder with a small network (3 units) of nano-satellites to be launched in mid 2020.
We will show the potential and prospects for short and medium-term development of the project, 
demonstrating the disrupting possibilities for scientific investigations provided by the innovative 
concept of a new ``modular astronomy'' with nano-satellites (e.g. low developing costs, very short realization time). 
Finally, we will illustrate the characteristics of the HERMES Technological Pathfinder project, 
demonstrating how the scientific goals discussed are actually already reachable with the first 
nano-satellites of this constellation. The detector architecture will be described in detail, showing 
that the new generation of scintillators (e.g. GAGG:Ce) coupled with very performing 
Silicon Drift Detectors (SDD) and low noise Front-End-Electronics (FEE) are  
able to extend down to few keV the sensitivity band of the detector. The technical solutions for FEE, 
Back-End-Electronics (BEE) and Data Handling will be also described.
\end{abstract}

\end{frontmatter}

\section{Introduction}
Gamma-Ray Bursts (GRBs) are one of the most intriguing and challenging phenomena 
for modern science. Their study is of very high interest for several fields of astrophysics, such as 
the physics of 
matter in extreme conditions and black holes, cosmology, 
fundamental physics and the mechanisms of gravitational wave signal production,
because of their huge luminosities, up to more than 10$^{52}$ erg/s, 
their red-shift distribution extending from $z\sim0.01$ up to $z > 9$ (i.e., much above that of 
supernovae of the Ia class and galaxy clusters), and their association with peculiar 
core-collapse supernovae and with neutron star/black hole mergers. 

Since their discovery, GRBs were promptly identified as having a 
non-terrestrial  
origin \cite{grb:1973}. 
First observations were done using radiation monitors onboard the VELA spacecraft constellation, 
that was a network of satellites designed to monitor atmospheric nuclear tests.
Between 1963 and 1970 a total of 12 satellites were launched and the constellation was operating until 1985,
with more sensitive detectors on later satellites.
By analyzing the different arrival times of the $\gamma$-ray photon bursts as detected by different satellites, 
placed in different locations around the Earth, it was possible to roughly estimate the direction of the GRB, later improved using additional and better detectors, reaching a precision of $\sim$10$^\circ$.
With a very similar approach, the Inter-Planetary Network (IPN\footnote{\url{https://heasarc.gsfc.nasa.gov/w3browse/all/ipngrb.html}},  including all satellites with GRB-sensitive instruments on-board) was organised by GRB scientists in late '70s,
aiming to localize GRBs for the observation of counterparts at other wavelengths.
Basing on the availability of operating instruments, the IPN in its lifetime has involved up to more than 20 different spacecrafts.
This experience demonstrates that the localization accuracy of GRBs is improved 
by increasing the spacing between different detectors, and also by a more accurate detector timing resolution.
The IPN localizations are usually provided in few days, and although can reach angular resolutions of arcminutes and often arcseconds, 
the current typical accuracy, at high energies, is of the order of few degrees. This was demonstrated, e.g., in the case of the discovery of Gravitational Wave (GW) electromagnetic counterparts \cite{GW:2017}.
Such huge error box is too large to be efficiently surveyed at optical wavelenghts, where
tens/hundreds of optical transient sources are usually found, increasing enormously the 
probability to find spurious correlations.
The best strategy here is to perform a prompt search for transients at high energies,  with a localization accuracy of arcminutes or arcseconds,
reducing the probability of chance association.

\section{HERMES Mission Concept}

The \emph{High Energy Modular Ensemble of Satellites} (HERMES) project aims to 
realize a new generation instrument for the observations of high-energy transients. 
The proposed approach here differs from the conventional idea 
to build increasingly larger and expensive instruments.
The basic HERMES philosopy is to realize innovative, distributed and modular instruments
composed by tens/hundreds of simple units, cheaper and with a limited development time.
The present nanosatellite (e.g. CubeSats) technologies demonstrates that off-the-shelf components for space use
can offer solid readiness at a limited cost.
For scientific applications, the physical dimension of a single detector should to be compatible with the nanosatellite 
structure (e.g. 1U CubeSat of 10$\times$10$\times$10 cm$^3$).
Therefore, the single HERMES detector is of course underperforming (i.e. it has a low effective area),
when compared with conventional operative transient monitors, but the lower costs and the distributed concept of the instrument demonstrate that is feasible 
to build an innovative instrument with unprecedented sensitivity.
The HERMES detector will have a sensitive area $>$50 cm$^2$, therefore with several tens/hundreds of such units 
a total sensitive area of the order of magnitude of $\sim$1~m$^2$ can be reached.

By measuring the time delay between different satellites, 
the localisation capability of the whole constellation is directly proportional to the number of components 
and inversely proportional to the average baseline between them. As a rough example, with a reasonable average 
baseline of $\sim$7000~km (comparable to the Earth radius, and a reasonable number for low-Earth satellites in suitable orbits) and $\sim$100 nanosatellites simultanoeusly detecting a transient, a source localisation accuracy of the order of magnitude of $\sim$10~arcsec\footnote{
$\sigma_\mathrm{pos} = \sigma_\mathrm{CCF}/Bc\sqrt{N(N-1-2)} \approx 10 \mbox{\,\,arcsec}$; where $B$ is the baseline, $N$ the number of satellites and
$\sigma_\mathrm{CCF}$ is the error associated with the cross-correlation function.}
can be reached, for transients with short time scale (ms) variability.

The current phase of the project, \emph{HERMES Technological Pathfinder} (TP), focuses on the realization of three
nanosatellites, ready for launch at mid-2020.
The purpose here is to demonstrate the feasibility of the HERMES concept, operating some 
units in orbit and to detect a few GRBs. The next phase of the project, \emph{HERMES Scientific Pathfinder} (SP), will demonstrate the feasibility of GRB localisation using up to 6--8 satellites in orbit.
Although in both these preliminary phases reduced ground segment capabilities will be used, i.e. reduced data-downloading with a few ground contacts/day, the complete development of the HERMES detectors is expected.
These activities will pave the way to the final HERMES constellation composed of hundreds of nanosatellites.  Detailed mission studies, including orbital configuration, attitude control strategy, and sensitive area distribution will be performed, as well as a proper planning of the ground segment allowing to reach the ambitious scientific requirements, i.e prompt diffusion of the transient accurate localization. Thanks to the production approach, the context of a typical Small or Medium-class space mission seems to be compatible with HERMES final constellation, where most of the resources will be devoted to the multiple launches and to the realization of the ground segment.

\section{Payload Description}

\begin{figure}[H]
\centering
\includegraphics[width=0.8\linewidth]{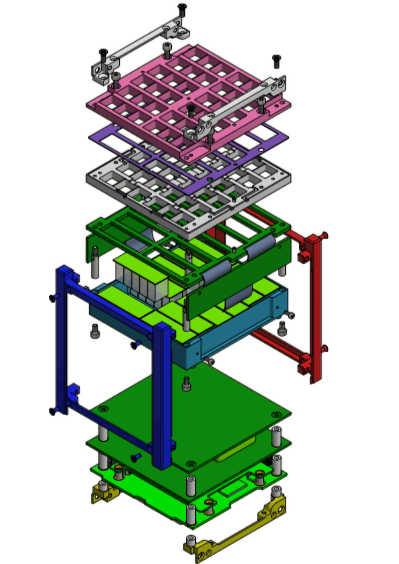}
\caption{Exploded view of the payload unit (10$\times$10$\times$10 cm$^3$) on board the HERMES nanosatellite.
From the top are shown the mechanical support composed by top (pink) and bottom (gray) parts, 
with optical filter (violet) in middle, the FEE board (dark green) allocating SDD matrices (light green), 
FE-LYRA chips on the top and BE-LYRA chips folded on the side (not shown for clarity), the GAGG crystal pixels 
(white trasparent) and their housing (greenish blue).
Mechanical ribs on top (grey) and on bottom (yellow) are also visible, necessary to fix 
the payload components to the satellite structure (blue and red).}
\label{fig:Payload}
\end{figure}

A possible solution for the HERMES payload is allocated in 1U-Cubesat 
(10$\times$10$\times$10 cm$^3$), cf. Figure \ref{fig:Payload}.
A mechanical support is placed on the instrument topside.
The support is composed by two parts to accomodate an optical/thermal filter in the middle.
The electronic boards for the Back-End and the Data Handling unit are
allocated on the bottom of the payload unit.
The detector core is located in the middle: this is a scintillator-based detector
in which Silicon Drift Detectors (SDD, \cite{Gatti:1984}) are used to both detect soft X-rays (by direct absorption in silicon)
and to simultaneously readout the scintillation light.
The payload unit is expected to allocate a detector with $>$50~cm$^2$ sensitive area
in the energy range from 3--5~keV up to 2~MeV, with a total power consumption $<$4~W
and total weight of $<$1.5~kg.

\subsection{Detector core architecture}

Aiming at designing a compact instrument with a very wide sensitivity band, 
the detector is based on the so-called ``siswich'' concept \cite{Marisaldi:2008, Marisaldi:2004}, 
exploiting the optical coupling of silicon detectors with inorganic scintillators.
The detector is composed by an array of scintillator pixels, 
optically insulated, read out by Silicon Drift Detectors.

In this concept 
the SDDs play the double role of 
read-out device for the optical signal from the scintillator and of an independent 
X-ray solid state detector. 
Low energy X-rays are directly absorbed by the SDD, 
while higher energy X-rays and $\gamma$-rays are absorbed in the crystal and the optical scintillation 
photons are collected by the same detector.
Only very low noise readout sensors and front-end electronics allow to reach a low energy scintillator threshold below 20--30 keV.
Above these energies the increasing sensitivity of the scintillator is able to compensate 
the lack of efficiency of thin silicon sensors (450~$\mu$m), so a quite flat efficiency in a wide energy band 
for the whole integrated system is reached.

The inorganic scintillators selected for this innovative detector is the recently developed \cite{Kamada:2012} Cerium-doped 
Gadolinium-Aluminum-Gallium Garnet (Ce:GAGG), a very promising material with all the required characteristics, i.e. a high light output ($\sim$50,000 ph/MeV), no internal radioactive background, no hygroscopicity, a fast radiation decay time 
of $\sim$90~ns, a high density (6.63 g/cm$^3$), a peak light emission at 520 nm and an effective 
mean atomic number of 54.4. All these characteristics make this material very suitable 
for the HERMES application.
Since GAGG is a relatively new material, it has not yet extensively investigated with respect 
to radiation resistance and performance after irradiation, although the published results are 
very encouraging \cite{Sakano:2014, Yanagida:2014, Yoneyama:2018}.
These tests showed that GAGG has a very good performance, compared to other scintillator materials largely used in the 
recent years in space-borne experiments for $\gamma$-ray astronomy (e.g. BGO or CsI), i.e. a very low activation background 
(down to 2 orders of magnitude lower than BGO), and a minor light output degradation with accumulated dose.

The SDD development builds on the state-of-the-art results achieved within the framework of the Italian ReDSoX
collaboration, with the combined design and manufacturing technology coming by a strong synergy 
between INFN-Trieste and Fondazione Bruno Kessler (FBK, Trento), in which both INFN and FBK co-fund the production of ReDSoX Silicon sensors.
A custom geometry for a SDD matrix (Figure \ref{fig:fee}) was designed, in which a single crystal 
($\sim12.1\times6.94$ mm$^2$) is coupled with two SDD channels.
Therefore, the scintillator light uniformly illuminates two cells, giving rise to a comparable signal 
output for both channels. This allows to discriminate scintillator events (higher energy $\gamma$-rays) 
by their multiplicity: lower energy X-rays, directly absorbed in the SDD, are read out by only one channel.

\subsection{Readout ASIC: from VEGA to LYRA}
\begin{figure}
\centering
\includegraphics[width=0.99\linewidth]{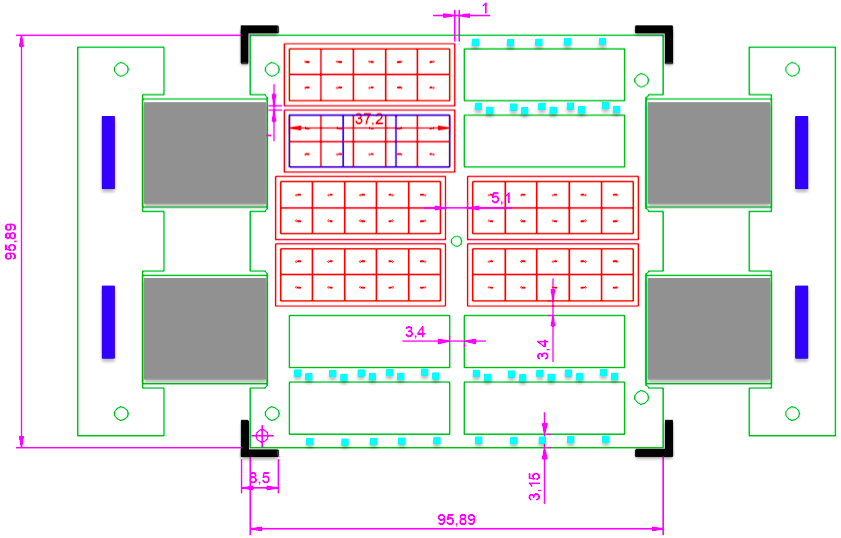}
\caption{Sketch of the top view of FEE board for the HERMES nanosatellite.
The black corner indicate the overall nanosatellite structure (10$\times$10 cm$^2$).
The board will allocate SDD matrices (in red) and FE-LYRA chips (light blue) very close to each SDD anode.
The BE-LYRA chips (blue) are allocated on the rigid part that will be folded on the side of the satellite, 
by means of flex cables (gray).}
\label{fig:fee}
\end{figure}

The HERMES detector, constituted by 120 SDD cells distributed over a total area of $\sim$92 cm$^2$, requires a peculiar 
architecture for the readout electronics. A low-noise, low-power Application Specific Integrated Circuit (ASIC) 
named LYRA has been conceived and designed for this task. LYRA has an heritage in the VEGA ASIC
 \cite{Mahdi:2014, Campana:2014} that was developed by Politecnico of Milano and University of Pavia 
within the ReDSoX Collaboration during the LOFT Phase-A study (ESA M3 Cosmic Vision program), although 
a specific and renewed design is necessary to comply with the different SDD specifications, 
the unique system architecture and the high signal dynamic range needed for HERMES. 
A single LYRA ASIC is conceived to operate as a constellation of 32+1 Integrated Circuit (IC) chips. 
The 32 Front-End ICs (FE-LYRA) include preamplifier, first shaping stage and signal line-transmitter, the single 
Back-End IC (BE-LYRA) is a 32-input ASIC including all the circuits to complete the signal processing chain: 
signal receiver, second shaping stage, discriminators, peak\&hold, control logic, configuration registers and multiplexer. 
The FE-LYRA ICs are small (0.9$\times$0.6 mm$^2$ die) allowing to be placed very close to the SDD anodes, 
in order to minimize the stray capacitances of the detector-preamplifier connection, maximizing the 
effective-to-geometric area ratio ($\sim$54~cm$^2$ vs. $\sim$92~cm$^2$). 
In this configuration (Figure \ref{fig:fee}), the BE-LYRA chips ($\sim$6.5$\times$2.5 mm$^2$ die) can be placed out of the detection plane, allocating SDD matrix and FE-LYRA ICs on a rigid part, by means of embedded flex cables. 
The flat cables allow also avoiding the additional space required by connectors, offering the possibility to ``fold'' the boards allocating the BE-LYRA chips (on a rigid part) at right angle with respect to the detection plane, on the external side of the payload unit.

\subsection{Back-End Electronics}
The Back-End electronics (BEE) of HERMES includes the BE-LYRA chips, external commercial analog-to-digital (ADC) converters
and a FPGA-based control logic. The control logic takes care of the signal handshaking 
required to read out analogue signals from BE-LYRA chips, syncronizing the digital conversion operations, 
and time tagging the events based 
on conventional GPS sensor, combining an atomic clock signal (CSAC) to reduce as much as possible the natural shift/jitter of the GPS sensor ensuring a sub-microsecond timing resolution.
Due to the peculiar architecture of the detector core, the BEE will also perform the 
Event Data Generator functionality, automatically discriminating the location of photon interaction 
(silicon or scintillator), on the basis of the multiplicity of the readout signals.
This fundamental task has to be carried out in real-time to generate the photon lists that include channel address, time of arrival of photons
and a raw energy estimation, which are mandatory for scientific data processing 
based on a suitable on-board logic.

\subsection{Payload Data Handling Unit}
The HERMES Payload Data Hadling Unit (PDHU) will be implemented on iOBC, manufactured by ISIS, a commercial on-board computer.
This model, with a weight of $\sim$100~g and an average power consumption of 400~mW, will implement all functionalities required for HERMES, 
such as telecommands (TCs), housekeeping (HKs), power system commanding (PSU), handling operative modes of the payload (by TCs or automatically),
generating the telemetry packets (TMs) and managing the interface with the spacecraft.
A custom algorithm making the satellites sensitive X-ray and $\gamma$-ray transients, continuously compare the 
current data rate of the instrument with the average background data rate taken previously.
When a transient occurs, the events, recorded on a circular buffer, are then sent to the ground on telemetry packets.
Due to the different families of GRB, ratemeters on different timescales, energy bands and different geometric regions 
of the detection plane will be implemented.

\section{Conclusion}
The HERMES project final aim is to realize a new generation instrument composed by hundreds of detectors onboard nanosatellites.
This disruptive technology approach, although based on ``underperforming'' individual units, allows to reach overall sensitive areas of the order of $\sim$1~m$^2$, with unprecedented scientific performance for the study of high-energy  transients such as GRBs and gravitational wave counterparts.
The current ongoing phase of the HERMES project (Technological Pathfinder), focuses on the realization of the three nanosatellites to be launched in mid-2020, that will demonstrate the proposed approach to detector design (Silicon Drift Detectors coupled to GAGG:Ce scintillator crystals) and its performance. 
In this framework, relevant prototyping activities are currently under development, towards the implementation phase.

\section*{Acknowledgments}
HERMES is a \emph{Progetto Premiale MIUR}.
The authors acknowledge INFN and FBK 
(ReDSoX2 project and FBK-INFN agreement 2015-03-06), ASI and INAF (agreements ASI-UNI-Ca 2016-13-U.O and ASI-INAF 2018-10-hh.0).

\end{document}